# Interstellar Grain Alignment – Observational Status

**B-G Andersson**

SOFIA Science Center,
Universities Space Research Association,
NASA Ames Research Center, M.S. N232-12
Moffett Field, CA 94035

**Abstract** Interstellar polarization in the optical/infrared has long been known to be due to asymmetrical dust grains aligned with the magnetic field and can potentially provide a resource effective way to probe both the topology and strength of the magnetic-field. However, to do so with confidence, the physics and variability of the alignment mechanisms must be quantitatively understood. The last 15 years has seen major advancements in both the theoretical and observational understanding of this problem. I here review the current state of the observational constraints on the grain alignment physics. While none of the three classes of proposed grain alignment theories: mechanical, paramagnetic relaxation and radiative alignment torque, can be viewed as having been empirically confirmed, the first two have failed some critical observational tests, whereas the latter has recently been given specific observational support and must now be viewed as the leading candidate.

## 1 Introduction

Interstellar optical polarization was discovered in 1949, independently by Hall (1949) and Hiltner (1949a,b). Already in Hiltner's second paper of that year the effect was ascribed to dichroic extinction by asymmetric dust grains aligned with the magnetic field. This is now well established and is supported (Figure 1) by comparing the amount of polarization with the visual extinction, which shows that the upper envelope of the former is linearly correlated with the lat-ter, and by comparing the position angle of the optical/near infrared polarization with the position angle of polarization due to synchrotron radiation, for which a well understood theory exists (e.g. Jackson 1975; Burke & Graham-Smith 1997)

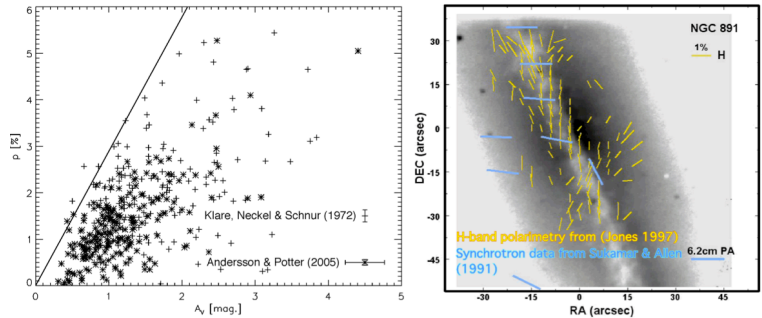

**Fig. 1.** That optical/near-infrared polarization is due to dichroic extinction by asymmetric dust grains aligned with the magnetic field is supported by the fact that (left) the upper envelope of the amount of polarization is correlated with the visual extinction (in this case towards a region around the Southern Coalsack; Andersson & Potter (2005)) and (right) that the position angles of the polarization agree with the position angles observed from synchrotron radiation (with an offset of ~90°, expected for zero pitch-angle electrons, cf Jackson 1975 ).

We can therefore safely assume that the optical/near-infrared (O/IR) polarization observed from the ISM is due to dichroic extinction by asymmetric dust grains aligned with (if not by!) the magnetic field. In the far-infrared (FIR), these aligned grains give rise to polarized emission which has been observed in a number of sources at wavelengths from the far infrared to mm-waves (cf. reviews by Hildebrand 1996 and summaries by Dotson et al. 2000, Vaillancourt & Matthews 2012, and references therein).

Significant progress has been made over the last two decades in both theory and observations and we are now at a point where quantitative comparisons between theory and observations are being made, and promises to reward us with a predictive, observationally supported theory of interstellar grain alignment within the foreseeable future.

This review will focus on the observational aspects of interstellar grain alignment physics and will only provide a brief summary of the theoretical aspects of the field. Excellent reviews of the latter can be found in e.g. Roberge (1996) and Lazarian (2007).

The main observational constraints on the grain alignment mechanisms come from (in the O/IR) the interstellar extinction and polarization curves (e.g. Fitzpatrick & Massa 1990; Cardelli, Clayton, & Mathis 1989, Serkowski 1973; Wilking et al. 1980). Based on the theory for scattering of light by small particles ("Mie theory" and its extensions; Mie 1908; Debye 1909; cf Whittet 2003; Krügel 2003; Draine 2011), complemented by elemental abundance constraints and estimates of the refractive indices for the grain materials, these allow an inversion from the wavelength dependence of the extinction and polarizization, respectively, to grains size distributions (by grain material) of the total grain population and the aligned part (Mathis, Rumple & Nordsieck 1977, Kim, Martin & Hendrey 1994, Clayton et al., 2003; Mathis 1986, Kim & Martin 1995). The dust is usually modeled as consisting of silicates, amorphous carbon and small graphite particles made up of simple shapes (e.g. spheres, spheroids or cylinders). The number density of the dust was found to be satisfactorily (Mathis, Rumple & Nordsieck 1977) modeled a power-law size distribution

$$n(a) \propto a^{-q}$$

with an exponent of $q \approx 3.5$, and grains at least in the range 0.005-0.25 mm range. Later models extend the upper limit with an exponential decay to larger grains (Kim, Martin & Hendrey 1994, Clayton et al., 2003, Draine & Li 2007; Draine & Fraisse 2009). These grain models are – of course – a simplification, which disregards such complications as possibly composite grains or silicate grains with carbonaceous mantles (Duley, Jones & Williams, 1989; Li & Greenberg 1997) or grains of highly complex shapes or structures. In the dark, cold parts of molecular clouds volatile ice mantles form on the refractory grains (Whittet 2003), which can be of substantial relative volume (Hough et al. 2008). In addition to their influence on the extinction and polarization properties of the material it is becoming increasingly clear that these ice mantles are crucial to the understanding of the chemical evolution of molecular clouds (Herbst & van Dishoeck 2009).

The polarized spectrum of MIR absorption and FIR/(sub)mm-wave emission from warm dust provide complementary constraints of the characteristics of the dust in terms of grain shape and refractive indices (e.g. Hildebrand & Dragaovan 1995, Draine & Fraisse 2009) for high column density line of sight.

The most direct probe of the grain alignment is through the fractional polarization ($p/A_V$). Although this, to first order, measures the fraction of aligned, asymmetric grains, it will, in detail, depend on a number of, often poorly constrained, parameters, including

1. The size distribution of the grains
2. The chemical composition and refractive index of the grains
3. The fraction of asymmetric grains
4. The alignment efficiency of the grains
5. The thermal and radiation field environment of the grains
6. The general orientation of the ordered magnetic field
7. The magnetic field line of sight topology, and level of turbulence

Therefore these factors, their uncertainties and – possibly – variations along each line of sight, need to be taken into account when comparing observations to theoretical predictions.

Polarization due to dichroic extinction by asymmetric grains aligned with the magnetic field, will only be sensitive to the plane-of-the-sky component component of the field. Hence variations in the line-of-sight to plane-of-the-sky ratio of the magnetic field will affect the O/IR fractional polarization. Since both Zeeman splitting (Verchuur 1969; Crutcher et al. 1975; Troland & Crutcher 2008) and Faraday rotation observations (Wolleben & Reich 2004 and ref. therein) probe the line-of-sight component of the field, direct "cross calibration" between these, nominally more quantitative, methods and O/IR polarimetry are inherently difficult (Andersson & Potter 2009)

To account for the maximum observed polarization at small to moderate opacities of $p/E_{B-V} \leq 9\ \%$ mag$^{-1}$ (Serkowski, Mathewson & Ford 1975) efficient alignment of the larger silicate grains is required (Jones 1996 and refs therein). Mathis (1986) could reproduce the wavelength dependence of the polarization assuming perfect alignment of the large silicate grains. Kim & Martin (1995), using a combination of extinction and polarization data, showed that even for small axis-ratios, oblate spheroidal silicate grains, combined with a population of spherical (silicate and carbonaceous) grains can reproduce the observed general maximum polarization fraction seen with a high alignment fraction.

Comparisons between the linear and circular polarization observed in the ISM indicates that the dust grains giving rise to the polarization are good dielectrics (Martin 1974, Martin & Angel 1976, Mathis 1986), consistent with silicate, but not with carbonaceous grains. Strong polarization has been detected in the 9.7 μm and 18 μm silicate spectral features (Smith et al. 2000), whereas the 3.4 μm aliphatic C-H stretch feature, usually associated with carbonaceous dust, does not show detectable polarization (Chiar et al. 2006). For these reasons it is often assumed that large carbonaceous are either absent, symmetrical or unaligned. The 2175 Å extinction "bump" shows associated polarization in 2 out of 30 lines of sight observed (Anderson et al., 1996; Martin, Clayton & Wolff, 1999). Since this extinction feature is usually thought to be due to small graphite grains it is possible that these can be aligned, at least in some environments, whereas the larger amorphous carbon grains required in some models (Clayton et al. 2003; Draine & Li 2007) might not be alignable.

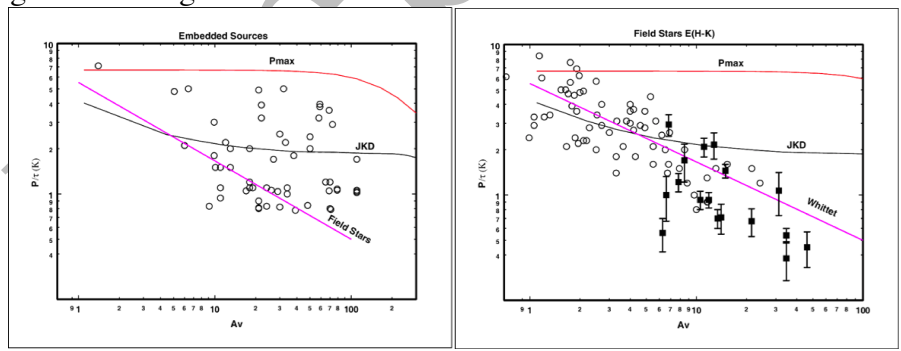

**Fig. 2** The fractional polarization in the K-band seen towards embedded sources and background field stars (Whittet et al 2008) are compared with the predictions from a model of polarimetric radiative transfer through a cloud with a turbulent magnetic field. Only at very large opacities without embedded sources does the fractional polarization drop below what can be explained by depolarization due to the random field component (addapted from Jones et al. 2011).

At large opacities the fractional polarization ($p/A_V$) drops systematically with opacity (Jones 1989, Jones, Klebe & Dickey 1992 (JKD); Goodman et al. 1992, 1995; Gerakines et al. 1995; Whittet et al., 2008), which is usually interpreted in terms of decreasing polarization efficiency. Some of this drop-off is, however, likely due to line-of-sight depolarization where different gas parcels in the cloud have intrinsic magnetic field directions (close to) perpendicular to each other and hence cancel out the polarization from each other.

This phenomenon has been studied by Meyers & Goodman (1991), Jones, Klebe & Dickey (1992), Ostriker, Stone & Gammie (2001) and Wiebe & Watson (2001). In general it is found that for multiple "decorrelation zones" and a large random field component a limiting behavior of $p \sim A_V^{1/2}$ is predicted, as is appropriate for a random walk process. Observationally, the level of turbulence, and hence the decrease in the fractional polarization that can be ascribed to a turbulent magnetic field, can be constrained by the fact that even for clouds with large opacities the observed polarization angles are not random (Jones 1996). For the Taurus cloud (observed by Whittet et al 2008; Figure 3) and even in the deep star-less cores (e.g. Crutcher et al. 2004) ordered magnetic fields are seen.

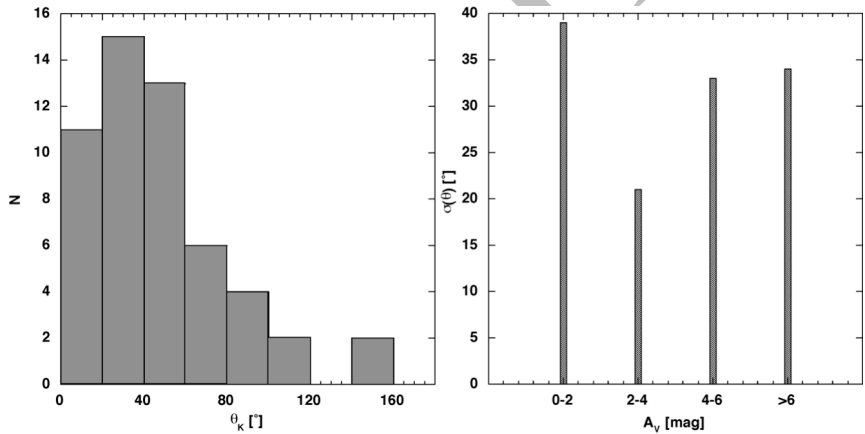

**Fig. 3** (Left) The distribution in polarization position angles for the "Field star" sample from Whittet et al (2008; Figure 2) shows a broad, but non-random distribution, indicating that the decrease in fractional polarization could not be fully ascribed to turbulence in the cloud. The dispersion in the position angles (right) shows no systematic variation with opacity.

Figure 2 shows the fractional polarization seen towards embedded and field star background sources (Whittet et al. 2008; Jones et al 2011) overlaid with the predictions from JKD using an equal mix of ordered and random components for the magnetic field with a decor-relation length for the random component of $\tau_K=0.1$ (Jones et al. 2011). For the lines of sight towards embedded sources, and for $A_V \leq 10$ for the field stars, the drop in fractional polarization can be accounted for by the random field's depolarizing effects, while towards even deeper sightline (with these field parameters, additional effects – likely a decrease in the grain alignment efficiency are required. Figure 3 shows the distribution of polarization angles observed for the field star sample of Whittet et al (2008). While the distribution is broad, it does not allow a depolarization fully dominated by magnetic field turbulence. Note, however, that this result should not be interpreted to mean that grain alignment variations are not present for $A_V<10$ mag, since the choice of the turbulence parameters is not well constrained in detail.

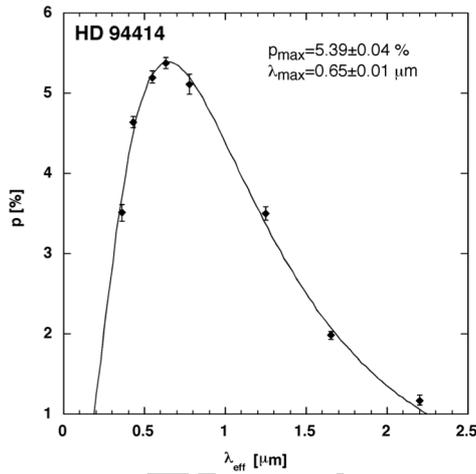

**Fig. 4.** The wavelength dependence of interstellar optical/near infrared polarization follows a universal relation referred to as the Serkowski relation (Equ. 1), here illustrated for the star HD 9441 behind the Chamaeleon I cloud. The data are from Whittet et al. (1992) and the fit uses equ. 2 to relate the K-parameter to $\lambda_{max}$.

The observed wavelength dependence for interstellar polarization shows a universal shape (Figure 4) with a maximum in the yellow-to-red range and falling off steeply both into the ultraviolet and infrared. The wavelength dependence can be parameterized (Serkowski 1973) by the relation:

$$p(\lambda) = p_{max} \cdot \exp\{- K \cdot \ln^2(\lambda_{max}/\lambda)\} \qquad (1)$$

known as the "Serkowski relation", where $p_{max}$ is the maximum amount of polarization seen at wavelength $\lambda_{max}$, and K controls the width of the curve. As was realized early on, based on Mie scattering theory (see Whittet 2003), the relatively narrow width of the polarization curve indicates that only a limited grain size distribution contributes to the polarization. More detailed modeling (Mathis 1986; Kim & Martin 1995) shows that the same size distribution can reproduce both the interstellar extinction and polarization curve, but with a larger small-size cut-off for the aligned grains around 0.04-0.05 µm.

Wilking et al. (1980, 1982) and Whittet et al. (1992) showed that K is correlated with $\lambda_{max}$ such that

$$K = (0.01 \pm 0.05) + (1.66 \pm 0.09) \cdot \lambda_{max} \qquad (2)$$

Andersson & Potter (2007) showed that a universal (at least for the 6 interstellar clouds in their study) relation exists between $\lambda_{max}$ and $A_V$ (Figure 5) of the form

$$\lambda_{max} = (0.166 \pm 0.003) \cdot <R_V> + (0.020 \pm 0.007) \cdot A_V$$

where $<R_V>$ is the average total-to-selective extinction in the cloud. Since the grain orientation randomization is assumed to be by gas-grain collisions and since these are more efficient for smaller grains, they interpreted this relationship as indicating radiatively driven grain alignment.

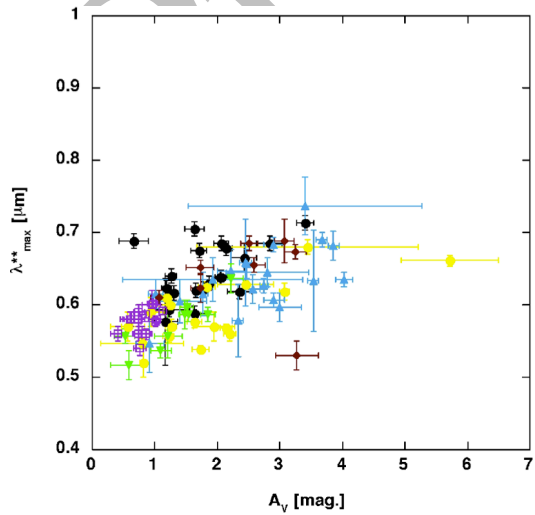

**Fig. 5** The wavelength of maximum polarization is seen to be linearly correlated with the visual extinction. Different colors correspond to data from different near-by clouds and the raw data have been adjusted to account of different average grains sizes (via $<R_V>$) and star formation rates on the different clouds. See Andersson & Potter (2007) for details.

Whittet & van Breda (1978), for a large sample of sightlines, found a correlation between $\lambda_{max}$ and the ratio of total-to-selective extinction, $R_V$, such that:

$$R_V = (5.6 \pm 0.3) \cdot \lambda_{max} \qquad ()$$

(cf. Whittet et al. (2001) and references therein), which is usually interpreted to indicate grain growth associated with larger values of $\lambda_{max}$, as $R_V$ is usually assumed to track the average grain size. Chini & Krügel (1983), however challenged this assumption and Whittet et al. (2001) noted that, for their Taurus sample, very little correlation was seen between $\lambda_{max}$, and $R_V$. Andersson & Potter (2007), expanded on the Whittet et al. results and argued that the derived variation of $\lambda_{max}$ with $R_V$ is due to the inclusion of different clouds in the same plot and argued that for any given cloud, in their 6 cloud sample, no clear relation is seen.

The FIR/(sub)mm-wave polarization spectrum shows a broad minimum around 350 µm for lines of sight in star forming cloud cores off of the FIR flux peak (Figure 6).

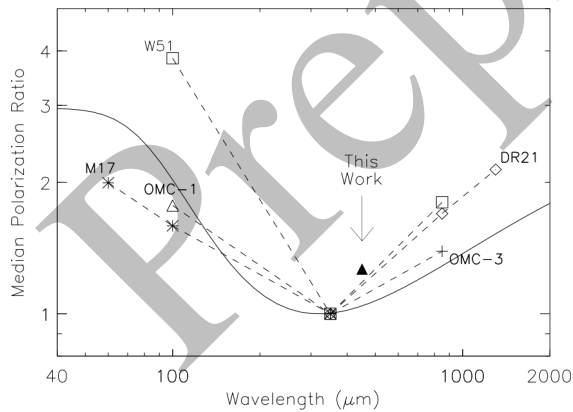

**Fig. 6** The observed polarization spectrum for a number of molecular cloud cores shows a systematic behavior with a marked minimum in the 350µm range. Overlaid (full drawn line) is an illustrative two-temperature dust model where only the hot component is polarized. (Reproduced from Vaillancourt et al. 2008 with permission from the AAS)

As shown, originally, by Hildebrand et al. (1999) a single population of grains cannot produce this polarization spectrum. At minimum, a two-component dust population is required with either significantly different grain emissivities or temperatures. Their preferred model, yielding the drop in polarization from 60 to 300 µm is one where the components differ in temperature and the warmer dust component is better aligned. As shown by Vaillancourt (2008) the up-turn in the polarization spectrum long ward of 350 mm can, in such a two-component model, be understood as an effect of different grain emissivities.

Hence, grain alignment is seen to vary with environment and in systematic ways, which should provide the observational constraints to allow an understanding of its physics.

## 2 Grain Alignment Theories

Three broad classes of grain alignment theories have been offered to explain ISM polarization: Mechanical alignment (Gold 1952a,b; Lazarian 1994, 1995, 1997a), paramagnetic relaxation alignment (Davis & Greenstein 1951; Jones & Spitzer 1967; Purcell 1979; Mathis 1986) and radiative alignment (Dolginov & Mytrophanov 1976; Draine & Weingartner 1996, 1997; Lazarian & Hoang 2007; Hoang & Lazarian 2008, 2009ab).

I will only briefly discuss these theories here and refer the interested reader to the above papers, the textbooks by Whittet (2003), Krügel (2003) and Draine (2011) and the review articles by Hildeband (1988), Roberge (1996) and Lazarian (2007) for details. While this discussion will focus on the mechanisms aligning the grains, it is crucial to also include the mechanisms randomizing the grain orientation when comparing theoretical predictions to observations. As shown by Draine & Lazarian (1998, their Figure 4), extrapolating to the grain sizes dominating the optical polarization; for neutral material gas-grain collisions are thought to dominate the grain disalignment. For the warm neutral and warm ionized media, infrared emission, and for dark clouds, plasma drag[1], may significantly contribute to the slowing of the grain rotation. Hence, variations in the gas density, temperature and ionization rate need to be considered when interpreting polarization data in terms of grain disalignment.

---

[1] Plasma drag is caused by the interaction of the grain electrical dipole moment with the surrounding ions.

For all alignment theories relying on the grain rotation becoming aligned with the magnetic field (i.e. paramagnetic and radiative alignment) the induced magnetization of a spinning grain, via the Barnett effect (see Draine 2011) assures that the grains have their axis of maximum inertia closely aligned with their angular momentum vector. The magnetization induced by the Barnett effect then causes the grain to Lamor precess about the external magnetic field. For thermally rotating grains at non-zero temperature, this alignment is not perfect (Lazarian & Roberge 1997) and becomes less so with rising grain temperatures, which affects the amount of polarization caused by the grains.

In **Paramagnetic relaxation** alignment, the grains are aligned with the external magnetic field due to the dissipation of internal magnetization energy. The dissipation occurs because the grain rotation makes the magnetic field - as seen by the grain material - vary rapidly enough that the material cannot fully respond to the variations in the applied field. This causes the magnetic susceptibility to become complex and leads to dissipation along those axes not parallel to the field.

The theory can, in turn, be divided into three sub-classes depending on whether (1) the grain magnetic susceptibility is typical for paramagnetic materials (e.g. silicates) and the grain is spun up by gas-grain collisions ("**Davis-Greenstein (DG) alignment**"); whether (2) the magnetic susceptibility is significantly enhanced over that of normal paramagnetic materials, as could result if the paramagnetic grain contained inclusions of ferromagnetic materials ("**Superparamagnetic alignment**"; Jones & Spitzer 1967; Mathis 1986; Lazarian 1997b); or whether (3) the rotation energy of the grain is well above the thermal energy of the surrounding gas ("**Suprathermal alignment**"; Purcell 1979), as may be the case if the grain rotation is driven by the ejection of photoelectrons or newly formed $H_2$ molecules. For any theory where the driving torques are fixed in the grain's coordinate system, a fundamental theoretical issue was raised by Lazarian & Draine (1999a,b). Because of the very strong nuclear Barnett relaxation, grains at T≠0, as large as 1 μm are likely to be "thermally trapped" in states with non-suprathermal energies, due to the continued, rapid flipping of the grain induced by the internal thermal excitations of the grain material. For such a grain the systematic torques never have the chance to bring the grain to suprathermal rotations speeds and only thermal alignment mechanisms are available. In all cases of paramagnetic alignment the grain is aligned with its angular momentum parallel to the magnetic field direction and hence the polarization traces out the direction of the projected field.

**Mechanical alignment** requires a systematic velocity difference between the gas and dust and assumes that gas collisions directly align the grains, somewhat like a water-wheel in a flowing stream. Such alignment results in a grain angular momentum perpendicular to the flow. The original mechanical alignment theory by Gold (1952a,b) assumed no magnetic field and considered the flow to be due to collisions between interstellar clouds oscillating around the galactic mid-plane as the source of the systematic flow motion. Incorporating magnetic fields, flows of charged particles can take place either along the field direction, as ambipolar diffusion across the field or as cyclotron orbits around the field lines. For mechanical alignment by (partially) ionized gas flows we would expect Gold alignment to produce polarization perpendicular to the magnetic field direction.

In recent modifications of the theory (Yan & Lazarian 2003) the magnetic field plays a crucial role also in the acceleration of the dust through a resonance between magneto-hydrodynamic waves and the cyclotron frequency of a charged grain. Since the relative motion of gas and dust grain is now perpendicular to the magnetic field lines, the predicted polarization is here along the projected field.

In mechanical alignment, which relies on collisions which make the grain tumble, the gas-dust flow must be supersonic, since - for subsonic flows - isotropic sound waves will tend to keep the grain orientation random (Gold 1952b; Lazarian 2007). Lazarian & Hoang (2007) have discussed mechanical alignment in which sub-sonic gas flows where gas collisions with helical grains generates grain spin with the angular momentum along the helicity axis. In parallel to the alignment by radiative torques where, because helicity is invariant under reflection, collisions from opposite sides spins a helical grain up in a common direction, this aligns the grains' spin axis parallel to the flow direction. In the presence of a dynamically im-portant magnetic field, the grains then align with the field and the polarization traces out the field direction.

The major advance in grain alignment theory of the last two decades is the quantitative development of **radiative alignment,** specifically through radiative alignment torque (RAT) theory. Originally proposed by Dolginov & Mytrophanov (1976), the theory proposes that grains with a net helicity can be spun up by the torques imparted by the radiation field as it is scattered off the grain. Because of the grain's helicity, the right- and left-hand circular polarization components of the radiation field have different scattering cross sections off of the grain and for an anisotropic field significant net angular momentum can be built up in the grain. As noted above, because helicity is invariant under reflection, thermal trapping is not a concern in radiative alignment and long duration spin-up can be achived for an anisotropic radiation field. As the grain acquires a magnetization via the Barnett effect (Purcell 1979) it precesses around the magnetic field and the continued radiative torques during this precession then aligns the grain with the magnetic field. Note that paramagnetic relaxation is not invoked in the alignment in RAT theory. As in the case of paramagnetic alignment, the grain is aligned with its angular momentum vector along the magnetic field and hence RAT alignment, generally, predicts polarization parallel to the magnetic field, as well. Draine and Weingartner (1996, 1997) used numerical calculations based on the DDSCAT code (Purcell & Pennypacker 1973; Draine & Flatau 1994) to show that significant alignment was achievable for a number of grain shapes, many of which were not particularly "cork screw like" (see Figure 1 of Draine & Weingartner 1997). Lazarian & Hoang (2007; LH07) proposed an analytical model, consisting of a spheroidal grain with an angled, offset, mirror which they showed could closely reproduce the numerical results from Draine & Weigartner. The generality of the analytical model of LH07 has allowed them to extend the theory and provide several specific theoretical predictions.

## *2.1 Theoretical Predictions Amenable to Observational Tests*

### 2.1.1 Paramagnetic Relaxation

2.1.1.1 Davis-Greenstein Alignment

For dust both spun up and randomized by gas-grain collisions it can be shown (e.g. Draine 2011) that the degree of alignment should *decrease* with grain size. The overall grain size distribution can be derived from the interstellar extinction curve (e.g. Mathis, Rumpl & Nordsieck 1977; Kim & Martin 1995; Clayton et al. 2003) while that of aligned grains can be determined from the polarization curve (e.g. Kim & Martin 1995).

Jones & Spitzer (1967) used thermodynamic arguments to show that significant alignment of grains with rotation driven by gas-grain collisions can only take place – at least for paramagnetic materials located in magnetic fields with reasonable ISM strengths –if the gas temperature is significantly different from the dust temperature. Since the gas heating at small opacities is dominated by collisions with photoelectrons emitted from small grains (e.g. Hollenbach et al 2009), and the cooling is dominated by line emission, the gas temperature drops rapidly with increasing opacity. The dust cools through optically thin continuum radiation, so dust temperature varies more slowly with opacity. Comparing observation of FIR dust emission and the line radiation from molecular tracers (e.g. Hotzel et al. 2001), the dust and gas temperatures can be compared. A recent, comprehensive, physical/chemical model of the temperature variations in a dark cloud (Hollenbach et al. 2009) shows that even for a quite elevated radiation field, as the visual extinction reaches $A_V \sim 4$ mag the gas ($T_g$=22 K) and dust ($T_d$=15 K) temperatures rapidly approach each other and by $A_V \approx 10$ mag are within a degree of each other.

- *Are smaller grains better aligned than large ones?*
- *Are grain aligned in environments where the gas and dust temperatures are approximately equal?*

2.1.1.2 Super-paramagnetic Alignment

As pointed out by Jones & Spitzer (1967) the thermodynamical constraints on thermally spun up paramagnetic alignment could be overcome if the magnetic susceptibility could be raised by several orders of magnitude. They suggested that this could be accomplished by the inclusion of small ferromagnetic sub grains into the silicate grain bulk consisting of metallic iron, iron -oxides or -sulfates (Jones & Spitzer 1967, Mathis 1986). Mathis (1986) expanded on this suggestion and showed that the observed polarization curve could be reproduced using a standard over-all power-law grain size distribution (Mathis, Rumple & Nordsieck 1977) and with the simple assumption that a silicate grain is aligned if it contains at least one such superparamagnetic (SPM) inclusion. He could also reproduce the observed correlation of the K parameter with $\lambda_{max}$ (equ 3), by assuming that both are driven by an increase in the size of the grains that contains one such SPM inclusion, while keeping the grain size distribution function and its upper cut-off constant. While the observed K vs. $\lambda_{max}$ behavior could also be accomplished by raising the upper bound on the grain sizes, Mathis argued that this would result in a variable red-to-near infrared extinction law, in contrast to what is observed.

It should be noted, however, that even for super-paramagnetic alignment, Roberge (1996) has argued, based on observational constraints from Hildebrand & Dragovan (1995) that even in the limit of infinitely large super-paramagnetic aligning torques the gas temperature must still be at least $5 \cdot T_{dust}$ for thermally rotating super-paramagnetic grains to be aligned.

- *Do the grains contain super-paramagnetic (SPM) inclusions?*
- *Do the ISM depletion patterns support the existence of grains with ferromagnetic inclusions?*
    o *If so, are grains with more Fe-rich materials better aligned?*
- *Is the K vs. $\lambda_{max}$ relation based on SPM inclusions unique?*

2.1.1.3 Suprathermal Rotation Alignment

Suprathermal spin-up ("Purcell") alignment requires that energetic particles are ejected from the grain surface in a way that yields a systematic torque. Thus the ejection sites must be localized on the grain surface over times long compared to the gas-grain collisional damping (Lazarian 1995). Purcell discussed three possibilities; inelastic collision with the gas particles, or the ejection of photo electrons or newly formed $H_2$ molecules. Of these $H_2$ formation provides the most promising mechanism. Lazarian (1995) revisited the mechanism, discussing the longevity of the sites of $H_2$ formation, including the effect of "poisoning" of the active surface sites by oxygen atoms. He estimated that Purcell alignment could be efficient for cold dust grains in material with significant atomic hydrogen available. For $T_d>20$ K he finds that the mobility of oxygen atoms leads to rapid poisoning of the active, chemisorption, sites and hence to short lived spin-up. However, laboratory experiments of $H_2$ formation (Pirronello et al. 1997, 1999), indicate that the formation under cold interstellar conditions take place through physisorbed particles and is efficient only at 6-10 K for olivine and 13-17 K for amorphous carbon (Katz et al 1999). Cazaux & Tielens (2004) argued that at higher temperatures the H atoms can access chemisorption sites, which would extend the $H_2$ formation to higher temperature (cf Cuppen, Morata & Herbst 2005). Whether the physisorption can be localized enough to allow for long-lived "Purcell rockets" is not clear.

In Lazarian's calculations the grain rotation speed drops to thermal values once $n_H/n=10^{-3}$. In most situations in the ISM we expect the medium to be in a state of detailed balance and hence any formation of $H_2$ molecules is associated with balancing destructions. Direct photodissociation of $H_2$ requires a photon of at least 14.5 eV and since this is shortward of the Lyman limit, no such photons are expected to be available in the general ISM. Destruction of the molecule is instead thought to take place through a two-step process initiated by the photoexcitation of the molecule's electronic structure (Field, Sumerville & Dressler 1966). On the subsequent relaxation into the ground electronic state, the molecule will end up in a vibrational state of v=14 or higher about 15% of the time which leads to vibrational dissociation (Draine 2011). The initial excitation into the Lyman band requires a photon short ward of 1108 Å. Assuming a standard interstellar extinction curve for $R_V$=3.1, $\tau(1108\text{Å})=1$ is reached already at $A_V \approx 0.24$ mag. The opacity for $H_2$ photodestruction is, in addition, strongly enhanced by the self-shielding in the lines, which causes the transition to $H_2$ to be very rapid (Federman, Glassgold & Kwan 1979; Viala 1986; van Dishoeck & Black 1986).

While the work function for interstellar grains is not well constrained, Weingartner & Jordan (2008) have estimated a value of 8 eV, corresponding to a photon of $\lambda \approx 1550$ Å. The $\tau(1550\text{Å})=1$ surface, under the same assumptions, occur at $A_V \approx 0.4$ mag.

Hence, unless the cloud material is highly porous, Purcell alignment should not be active beyond about 1 magnitude of visual extinction. Sorrell (1995) proposed that cosmic ray driven evaporation of grain mantles could drive grains at deeper levels to suprathermal spin velocities. However, Lazarian & Roberge (1997) showed that for this mechanism to work the cosmic ray fluxes has to be 6-7 orders of magnitude larger than expected for molecular clouds. Also, as noted above, the discovery of the nuclear Barnett effect (Lazarian & Draine 1999a,b) adds significant doubt to the ability of most interstellar grains to reach suprathermal speeds through a mechanism fixed in the grain's coordinate system because of the effect of thermal trapping where a slowly rotating grain.

– *Is grain alignment seen in environments where the high-energy radiation has been excluded?*
– *Is molecular hydrogen formation associated with enhanced polarization?*

### 2.1.2 Mechanical Alignment

For "classical" mechanical alignment, where the relative gas-dust flow is along the magnetic field lines, the most distinctive prediction of the theory is that the grains will be aligned with their long axis along the field direction and hence cause FIR polarization parallel to it (or perpendicular to the field for O/NIR dichroic extinction). For regions where the magnetic field orientation can be deduced independently and where varying line-of-sight opacities are not an issue, a 90° rotation of the polarization over small scales would indicate a change in alignment mechanism.

As discussed in Li and Houde (2008, and references therein) the velocity dispersion cut-off on small scales in molecular clouds may be used as a probe of the ambipolar diffusing in the material. It should thus, in principle, be possible to probe the suggestion by Lazarian & Hoang (2007) of ambipolar-driven mechanical grain alignment. Whether it can separated from other alignment variations and line-of-sight effects remains to be seen

– *Is polarization perpendicular to the magnetic field direction observed?*
– *Is the fractional polarization in molecular cloud correlated with the amount of ambipolar diffusion?*

**2.1.3 Radiative Alignment**

RAT alignment predicts that for grain with the required helicity, the radiation field will couple efficiently and, for an anisotropic radiation field, produce alignment as long as $\lambda<2a$, where $\lambda$ is the wavelength of the radiation and $a$ is the radius of the grain. This means that a broad range of grain sizes can be aligned under most interstellar conditions, and yields a number of testable predictions.

The size distribution of aligned grains (and hence the polarization curve) can potentially be understood by this simple condition (that grains are aligned for $\lambda<2a$), as well as some of the observed "anomalies" of interstellar polarization.

In the grain size distributions derived based on maximum entropy method inversions, Kim & Martin (1995) find a smallest aligned grain with a size of $a\approx 0.04$-$0.05\mu m$, half of the wavelength of the Lyman limit $\lambda=0.0912$ $\mu m$. Hence the grain alignment on the small size is potentially simply due to the lack of radiation short ward of the Lyman limit in the general ISM. A closely parallel argument to that of Mathis (1986) can then be made to explain, also under RAT alignment, the correlation between K and $\lambda_{max}$ in the Serkowski formula. Namely, if the underlying grain size distribution (both total and the asymmetrical fraction) is fixed and the smallest aligned grain is determined by the wavelength cut-off of the available light, then the polarization curve will narrow (K increase) and shift ($\lambda_{max}$) to the red, with a reddened radiation field, similarly to the Mathis model.

As the remaining radiation field becomes increasingly reddened, the condition $\lambda<2a$ should eventually no longer be fulfilled, even for the largest grains. The large cut-off in the grain size distribution is only weakly constrained by observations (by NIR extinction and total elemental abundances; Mathis 1986, Kim, Martin & Hendry 1994; Clayton et al 2003), but a steep drop-off in the mass distribution of the silicate grains is usually predicted in the 1-2 μm range. Using the interstellar extinction curve, together with the condition $\lambda<2a$, to calculate the opacity at which the largest grains are no longer aligned, we find, depending on the value of the total-to-selective extinction and upper grain size limit, of $A_V \approx 7$-$24$ mag.

Since - for RAT alignment - both the grain heating and the alignment is driven by the radiation field, a correlation between grain temperature and alignment is expected, both at small opacities, driven by the interstellar radiation field – or nearby stars – and in cloud cores, heated by embedded YSOs.

As noted by Hoang & Lazarian (2009) RAT alignment can take advantage of other mechanisms to enhance the alignment efficiency and fraction. Both superparamagnetic grains and suprathermal rotation can be incorporated into the RAT paradigm. "Pin-wheel torques" ($H_2$ or $e^-$ ejections) can lift grains out of low angular momentum states and enhance the alignment. For radiatively aligned superparamagnetic grains the alignment should be perfect and hence any tertiary alignment drivers (e.g. pinwheel torques) should not be able to further improve the fractional polarization.

A likely unique prediction of RAT theory is that the alignment should vary as the angle between the radiation and magnetic fields (LH07). The exact dependence and sign of the variation depends on the grain characteristics and color of the radiation field, but for "typical" grains in the diffuse galactic field, LH07 predicts enhanced alignment when the radiation field anisotropy is parallel to the magnetic field direction.

- *Does the alignment vary with the color of the radiation field?*
    - *Due to reddening*
    - *Due to the intrinsic SED of the radiation field?*

- *Are the opacities of the "polarization holes" seen in star-less cores consistent with the upper grain size limits derived from extinction and elemental abundance observations?*
- *Can FIR/sub-mm wave polarization from star forming clouds be accounted for?*
- *Does the grain alignment efficiency vary with the angle between the radiation and magnetic fields?*
- *Does $H_2$ formation enhance the fractional polarization?*

## 3 Observational Considerations

As with any tracer of low to moderate optical depth, polarization measurements are prone to uncertainties due to line-of-sight variations in space density and temperature of both the dust and gas, as well as in other parameters such as the ionization fraction, grain charge and radiation field. The gas-grain collision rate depends on the gas density and temperature, which disaligns the grains, but may also - for paramagnetic alignment - generate the alignment. For most tracers of these entities, the characteristic depth of the observations (i.e. the depth into the cloud dominating the measurement) is generally not well determined and matching gas and dust temperature to a given gas parcel is difficult. Whereas emission line observations, for instance CO (J=1-0), are dominated by the $\tau=1$ surface of the emission line, the opacity of the continuum radiation from the dust is much smaller and usually probes the full line of sight. In a few cases specific (limiting) opacities can be reliably directly sampled. The use of ice mantles on the dust grains provide one such case, where their existence can be used to derive a minimum opacity and tight temperature constraints on the gas and dust (Hollenbach 2009)

If the grain alignment falls with increasing opacity - and assuming that the material is in a gravitationally bound cloud with density increasing toward the center - the line of sight weighting of an observation is complicated. As is the case with other observations of diffuse material, mapping of a cloud can provide the means to a nominal "inversion" of the observations into depth dependence of the characteristics of the gas and dust (e.g. Jones, Klebe & Dickey 1992; Whittet et al 2008), but the sampling density of such maps for optical/NIR polarimetry depends on the density of suitable background sources, which can be a significant problem for high opacity regions with the limited-size telescopes usually available for polarimetric observations. As the ISM extinction curve rises towards the blue, increasingly long wavelength bands need to be employed for larger opacities. However, beyond about K-band (2 μm) the thermal emission (and opacity) of the Earth's atmosphere means that the number of sources bright enough to allow high quality polarimetry becomes much smaller than at shorter wavelengths. FIR/(sub)mm-wave polarization provides a way around the necessity for suitable background targets, but adds the complication of emissivity and dust temperature variations along the line of sight, and has only recently become sensitive enough to probe more than the brightest regions.

In particular if the alignment is directly driven by the radiation field (suprathermal or RAT alignment), it is important to note that the line-of-sight opacity (e.g. $A_V$) is not necessarily a good measure of the radiation field experienced by the grain. This can be due to the geometry of the cloud (i.e. a sightline in the outskirt of a prolate cloud pointing towards the observer will show a significantly higher $A_V$ than the opacity to the ISRF seen by a grain in that line of sight) or because of near-by bright radiation sources (Figure 7).

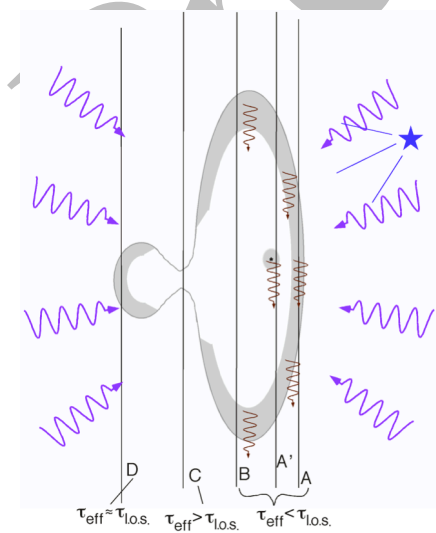

**Fig. 7** The line of sight (l.o.s.) opacity might not always be a good tracer of the radiation field that a dust grain sees, for non-spherical clouds or clumpy materials. Also, if there are stars near enough to the cloud to be of comparable strength to the diffuse radiation field (purple) the use of $A_V$ as indicator of the relative illumination of the dust can lead one astray, (Adapted from Andersson & Potter 2007, with permission from the AAS)

Andersson & Potter (2007) discussed various such observational biases and showed e.g. that most of the outliers in plots of $\lambda_{max}$ vs. $A_V$ could be accounted for by them. The differences in the fractional polarization variations with opacity for background field stars and embedded sources (Whittet et al 2008) are also likely due to the additional radiation field experienced by the dust from the embedded sources. For FIR polarimetry, internal radiation sources (Vaillancourt 2002; Vaillancourt et al. 2012) impose similar considerations and must be accounted for when analyzing the polarization in terms of grain alignment. Given these variations in environmental conditions, combining small sample polarization studies from different regions must be done with care.

In addition to these "scalar" line of sight issues, polarization measurements also depend on the topology of the magnetic field. That is, if the inherent polarization *orientation* of successive parcels along the line of sight of the material varies, the total polarization will depend not only on the relative grain alignment, but also on that relative orientation of the field. As discussed in the introduction, the consequences of such variations in the field topology can be difficult to disentangle from variations in the grain alignment efficiency. Detailed modeling of observations incorporating both stochastic magnetic field and alignment efficiency need to be done to evaluate the relative contributions to the loss of fractional polarization observed.

One way of avoiding the uncertainties associated with the depolarization effects, albeit at the cost of requiring observations also of short wavelength (UBV) polarization, is to use the wavelength dependence of the polarization, rather than the (fractional) amount of polarization observed. For optical polarization the Serkowski curve can be shown to trace the characteristic size of the aligned grains. As shown by Kim & Martin(1995) the location of $\lambda_{max}$ is most sensitive to the smallest aligned grains. Since these are also the grains most easily disaligned by collisions, we can use this measure as a grain alignment tracer.

# 4 Observational Constraints on Grain Alignment Mechanisms

## *4.1 Davis-Greenstein Alignment*

– *Large grains are better aligned than small ones*

Comparing the grain size distribution for all grains to that for the aligned grains required to produce the observed shape and location of the polarization curve and the extinction curve (e.g. Kim & Martin 1995), shows that the large grains (a~0.1 μm) are relatively much better aligned than the smaller ones (a~0.01 μm), in contradiction to the prediction of Davis-Greenstein theory.

– *Grains are aligned also in environments ($A_V$>5) where $T_{gas} \approx T_{dust}$*

Combining H-band polarimetry with FIR continuum observations and CO (J=1-0) data, Jones, Hyland & Bailey (1984) probed the star-less dark core "Tapia's Globule 2" in the Southern Coalsack ($A_V$~12 mag), and found that, although the derived dust and gas temperatures were almost identical at around 10 K, the dust in the core is, at least partially, aligned. Magnetic field estimates using the Chadrasekhar-Fermi method (Lada et al. 2004) yield B ≈ 25 μG, much less the ~1 mG that would be required to align the grains using Davis-Greenstein alignment (Jones, Hyland & Bailey 1984)

A very direct probe of the grain alignment at large opacities is provided by observing the relative polarization seen in solid state spectral features of molecules frozen out onto the grain surfaces. For several ices detected in the ISM, threshold extinctions exist below which no solid state features are detected (Whittet et al 1988, 1989; Williams, Hartquist & Whittet 1992). This threshold extinction varies both by species and from cloud to cloud, but is typically about $A_V \approx 3$ mag for water ice (observable at λ=3.1 μm) and $A_V \approx 6$ mag for CO ice (observable at λ=4.7 μm) and thus trace material deep into the clouds where the gas and dust temperatures approach each other. The detection of a significant polarization component in the CO ice line towards background stars, particularly in clouds with no, or only low-mass, star formation thus directly shows that grains are aligned at a depth into the cloud where the gas and grain temperature approach each other. Observations by Chrysostomou et al. (1996) towards W33A and, particularly, Hough et al (2008) towards the background star Elias 3-16 probing the Taurus cloud (Figure 8) show strong polarization in the CO ice line and thus put severe doubt on thermal paramagnetic alignment.

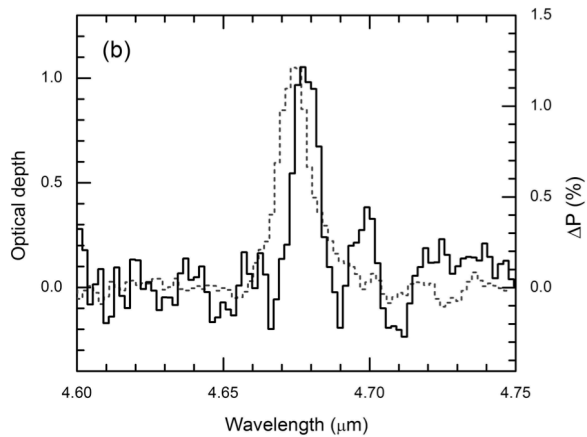

**Fig. 8** The CO ice line towards the background star Elias 3-16 shows significant polarization, indicating that grains are aligned at $A_V > 6$ mag. The dashed line (left-hand scale) shows the opacity while the full line (right hand scale) shows the polarization, with the continuum removed, (Reproduced, with permission from Hough et al. 2008)

*4.2 Super-paramagnetic Alignment*

– *Super-paramagnetic inclusions are seen in Interplanetary Dust Particles (IDPs)*

A way to overcome the predicted relatively better alignment efficiency for small grains in thermally spun up grains under paramagnetic alignment was put forward by Mathis (1986), who proposed that a dust grain will be aligned if, and only if, it contains at least one SPM inclusion. Using dust parameters derived from extinction data and the above assumption he could reproduce the polarization curve for $\lambda_{max} = 0.55$ μm, assuming a cut-off in the aligned grains at $a' > 0.09$ μm. Using estimates of the enhancement of the paramagnetism from Jones & Spitzer (1967) he derived a characteristic size of the SMP inclusions of 0.01 μm. Adding UV polarimetry to the constraints, Wolff, Clayton & Meade (1993) modified the cut-off of the aligned grains somewhat to about 0.06 < a < 0.1 µm. While there is some doubt about how such grains are generated (Martin 1995), the identification of amorphous silicate grains - Glass with Embedded Metals and Sulfides (GEMS) - in interplanetary dust particles from comets (Bradley 1994), lent support to the hypothesis. Goodman & Whittet (1995) argued that the dark patches seen, using transmission electron microscopy, inside these GEMS containing Fe(Ni) metals and iron-rich sulfides, are of the right size and volume-filling factor to match the requirements of Mathis' theory and exhibit super-paramagnetic susceptibilities.

– *The fractional polarization is not correlated with the interstellar solid state iron not included in silicates*

Very recently, Voshchinnikov et al (2012) cast doubt on the generality of such SPMs in the ISM, by comparing the mineralogy of interstellar dust with its polarizing characteristics. They used an extensive survey of the depletion patterns of Si, Mg and Fe from Voshchinnikov & Henning (2010), combined with new and previously published polarization data to probe for a correlation between ferromagnetic compounds and enhanced polarization. Based on the depletion measurements they calculate the amount of solid phase iron available to make ferromagnetic materials, by assuming that the depleted Silicon and Magnesium are fully used to form minerals with the stoichiometry of Olivine.

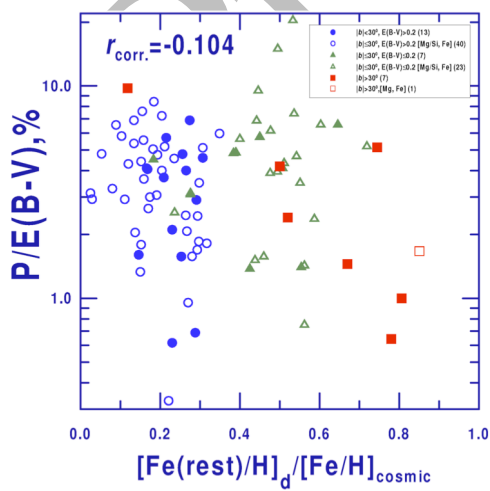

**Fig. 9** The fractional polarization is plotted against the iron in the solid phase, but not used up in silicates with olivine stoichiometry. If the assumption is made that this "remaining" solid iron is – at least partially – in the form of ferromagnetic compounds, and super-paramagnetic alignment is active, the fractional polarization should increase with the abundance of Fe[rest]. No such increase is seen (reproduced sith permission from Voshchinnickov et al 2012)

They then correlated the "remaining" solid phase iron, where various iron oxides are assumed to dominate, with the fractional polarization ($p/E_{B-V}$) for 95 lines of sight. No enhanced fractional polarization is seen (Figure 9).

### *4.3 Suprathermal Spin Alignment*

– *Grain alignment occurs at opacities where the short wavelength radiation has been excluded*

Whittet et al (2008) assembled fractional polarization measurements for the Taurus cloud from its surface to about $A_V=20$ mag. and have shown that a single power-law fit ($p/A_V \propto A_V^{-b}$) with an exponent of b = -0.52±0.07 can describe the full range for their field star sample. Andersson & Potter (2007) have shown that the wavelength of maximum polarization follows a universal linear relation with $A_V$ (once variations in the average value of $R_V$ between clouds is accounted for) to at least $A_V \approx$4-5 mag. In neither case are any changes seen in the relations at the opacities where the high-energy photons responsible for $H_2$ destruction of photoelectric emission have been excluded. It is, however, noteworthy that even though the observed transition from atomic to molecular gas is rapid, it is less abrupt than might be implied by line opacities. In a sample of 23 line of sight observed in $H_2$ with the Far Ultraviolet Spectroscopic Explorer with visual extinctions up to $A_V \approx 3.4$, Rachford et al (2002) found no cases of a molecular fraction f=$2N(H_2)/[2N(H_2)+N(H\ I)]$ > 0.8. Hence the transition to fully molecular material is significantly slower than might be expected and atomic hydrogen will be available also at opacities of the order a few.

– *$H_2$ formation enhanced grain alignment is observationally suggested in the reflection nebula IC 63*

Since the destruction of $H_2$ in the ISM takes place through a process of line absorption followed by a fluorescent cascade, and the time scale for $H_2$ formation-destruction cycling is short compared to the dynamical time scales of a photodissociation region (PDR) the NIR fluorescent emission from the molecule traces its formation rate. Andersson et al. (2012, submitted) used this argument to search for $H_2$ formation driven grain alignment in the reflection nebula IC 63. Figure 10 shows the measured polarization for the stars probing the nebula as a function of $H_2$ 1-0 S(1) emission. If allowance is made for the strongly enhanced collision rate associated with the bow shock region in the nebula (probed by target #46 in the Figure), a statistically significant correlation is seen between $H_2$ formation and alignment.

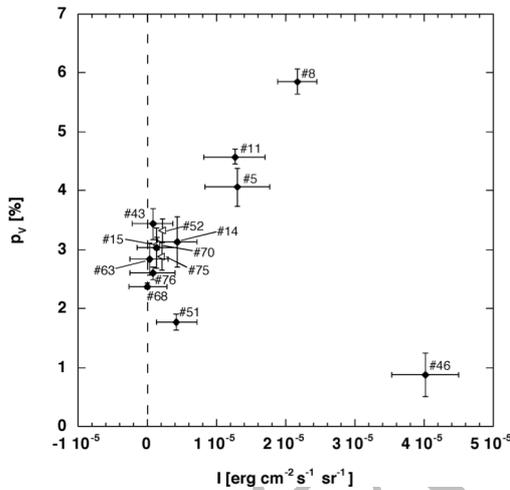

**Fig. 10** The amount of polarization towards stars behind the reflection nebula IC 63 is correlated with the fluorescent intensity in the $H_2$ 1-0 S(1) line. The fluorescence traces the $H_2$ destruction and, since the time scales for reformation on the dust grain surfaces is much shorter than the dynamical time scale of the PDR, also the $H_2$ formation rate. Target #46 is located behind the bow shock of the nebula where intense collisional disalignment dominates (Reproduced from Andersson et al. 2012, with permission from the AAS).

*4.4 Mechanical (Gold) Alignment*

– *In some circumstellar environments of high-mass YSOs the polarization seems twist by 90° close to the origin of the outflows.*

High spatial resolution observations of several high-mass young stellar objects (YSO) show complicated position angle structures in the innermost regions around the circumstellar disk and outflows. For instance, Rao et al (1998) studied the Kleinman-Low object in Orion; Cortes, Crutcher & Mathews (2006) observed NGC 2071 IR with BIMA and Tang et al. (2009) used the SMA to map G5.89-0.39 in polarized (sub)mm-wave emission. In all cases a (close to) 90° rotation of the polarization is observed when comparing the emission along and across the molecular outflow direction. It is still not clear whether this is better interpreted as evidence of very complicated field geometries or support for mechanical grain alignment. The discovery of synchrotron emission from the jet of a massive YSO (Carrasco-Gonzales et al 2010, 2011) promises to allow an independent tracer of the magnetic field structure and to thus conclusively probe the possibility of mechanical alignment in these sources. As discussed by e.g. Hildebrand et al (1999), opacity effects can cause polarization angle rotations by 90° with wavelength as the polarization changes over from being dominated by emission to absorption, as seen e.g. towards the Galactic center (Dowell 1997, Novak et al. 1997), but such effects are unlikely to be the cause of the effects seen in high-mass star forming regions.

*4.5 Radiative Alignment*

– *Grain alignment is active at large opacities and varies smoothly with opacity.*

As noted above (Sec. 4.3), the fractional polarization to optical depth much beyond $A_V=1$ mag. shows a smooth decline with a power-law index of about -0.5, much less than the -1 which would be expected if the polarization was produced only in a relatively thin skin. Similarly the location of $\lambda_{max}$ varies linearly with $A_V$ to – at least – $A_V=4-5$ mag.

– *The grain alignment efficiency is correlated with the dust temperature.*

While, as discussed above, the observations of grain alignment in regions where little or no temperature difference exists between the gas and the dust, puts severe doubts on DG alignment, direct variations is the dust temperature should be correlated with the alignment efficiency in RAT alignment since the photons that align he grains will also heat them. Matsumura et al. (2011) have used optical spectropolarimery to shown that the polarization efficiency is enhanced when the dust temperature is elevated for stars in the Pleiades cluster. A direct correlation between grain alignment and heating is also seen for the dust surrounding HD 97300 in Chamaeleon I (Anders-

son & Potter 2010)

As discussed above, the polarization spectra in the FIR/(sub)mm-wave regime (Figure 6) requires at least a two component dust distribution, where the warmer dust is better aligned. Whereas differences in the refractive indices might cause some differences in grain temperature, a more natural explanation, considering the observational data for the O/IR is that the warmer grain component is exposed to a stronger radiation field (Hildebrand et al 1999; Vaillancourt 2002, Vaillancourt et al. 2008)

- *The "polarizations holes" seen in deep starless clouds are consistent with the loss of alignment of the largest grains.*

At very high opacities in star-less cores a steep drop in the fractional polarization is seen in sub-mm wave polarimetry (e.g. Chrutcher et al. 2004) and hinted at in NIR data (Jones et al, 2011; Figure 2) where the slope in the fractional polarization with $A_V$ steepens from a value of about -0.5 (Whittet et al 2008) to about -1. Jones et al. (2011) have speculated that these "polarization holes" are simply the effect of that the extincted radiation eventually becoming too red to couple to any remaining grains.

- *The grain alignment efficiency depends on the angle between the radiation and magnetic fields*

Andersson & Potter (2010) and Andersson et al. (2011) tested the RAT theory prediction that the grains alignment should depend on the angle between the radiation and magnetic fields by observing the polarization in the Chamaeleon I cloud in the area surrounding the B9 V star HD 97300. Located about 0.3 pc from the near side of the cloud (Jones et al. 1985) the star strongly illuminates the underlying could. For the region of the cloud in which the star dominates the radiation field (d≈0.75°) a correlation is seen between grain heating and alignment efficiency.

We measured a number of stars at various position angles from the star relative to the magnetic field direction and detect enhanced grain alignment at a 9σ level in the magnetic field direction. We note that, as no position angle rotation was seen, this enhancement is not likely to be due to mechanical alignment. Fitting a toy-model of the grain heating assuming thin "pizza box shaped" grains we found a good fit to the I(60)/I(100) color temperature. The derived fraction of aligned grains is in close agreement with those derived by independent mean by Kim & Martin (1995).

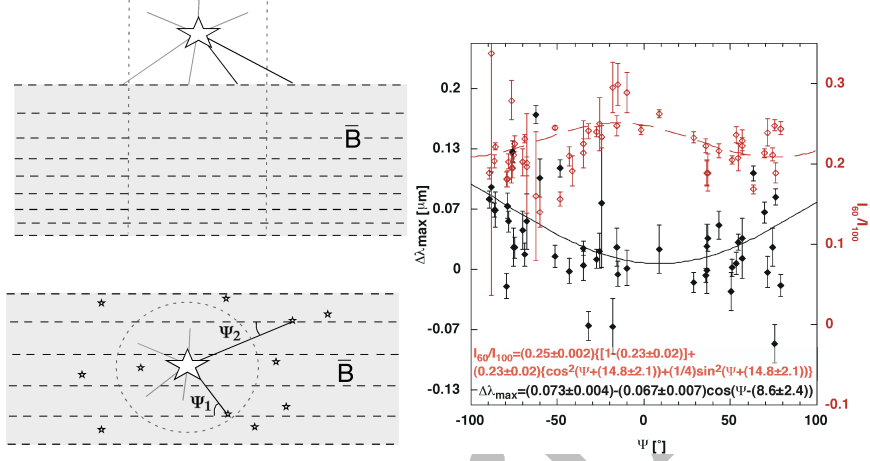

**Fig. 11** For a star close enough to an interstellar cloud to dominate the radiation field, RAT theory predicts that the alignment should vary with the angle between the magnetic and radiation fields. We can probe this by observing background targets located around the illuminating star (left). In accordance with predictions, Andersson et al (2011) found a maximum in the grain alignment (minimum in $\Delta\lambda_{max}$) at $\Psi=0$ (black symbols). The red symbols and curve show the color temperature and best fit to a toy model of the differential heating of the aligned grains (see Andersson et al. 2011 for detail).

## 5 Summary and Conclusions

Grain alignment is observed to be a universal phenomenon from the diffuse gas to molecular cloud cores (at least those containing star formation). It is seen at large opacities ($A_V$~5-20) - even in clouds without internal heat (or radiation) sources, where the gas and dust temperatures closely approach each other. Most uncontrovertibly this is illustrated by the alignment of the CO ice line towards Elias 3-16 in Taurus (Hough et al. 2008). This would seem to exclude thermal paramagnetic alignment, even for super-paramagnetic grains (Roberge 2004), particularly in light of Voshchinnickov et al.'s (2012) results that the amount of (presumably) ferromagnetic iron compounds in the dust dues not correlate with the grain alignment.

Combining the JKD/MG depolarization models with the drop-off in the fractional polarization shows that - while the grain alignment falls into the cloud, grains are aligned at large opacities. Using the wavelength of maximum polarization as the tracer of the grain alignment we find a linear relation with $A_V$ to at least $A_V$=10 and possibly deeper (Andersson et al 2012; in preparation). At these opacities, the energetic photons needed to dissociate the hydrogen molecules or cause photoelectric emission will be excluded. This makes also suprathermal alignment difficult to reconcile with observations for large parts of molecular clouds. However, $H_2$ formation enhanced alignment might have been observed in the reflection nebula IC 63.

For "classical" mechanical alignment where relative gas-dust flows are along the magnetic field lines, the observed polarization direction in the ISM generally are inconsistent with the alignment predictions. For several high-mass YSOs rapidly rotating polarization field in the inner part of the disk/outflow may however indicate mechanically driven alignment in these environment. For the more recently proposed versions of mechanical alignment (via ambipolar diffusion or through cyclotron resonant acceleration of charged dust grains) the theoretical predictions are either not yet clear enough to provide unambigous, unique predictions or these have not been tested.

In contrast, radiative alignment torque theory has - over the last decade and a half - partly in response to the difficulties identified for earlier theories, been developed into a self-consistent theory providing specific, testable, predictions. Some of these can explain hitherto puzzling behavior of the observed polarization, while some are new and require specific observations to be performed. The drop in the fractional polarization with depth and the shift $\lambda_{max}$ to longer wavelength with increasing $A_V$ is a natural consequence of the extinction and reddening of the diffuse Galactic light into the cloud and could also explain the observed K vs. $\lambda_{max}$ correlation. The relatively good polarization seen in the FIR/(sub)mm-wave for star forming clouds and the contrasting behavior between star forming and starless cores is also consistent with RAT alignment. The fact that grains are heated as well as better aligned next to bright (blue) star also speaks for radiative alignment (Andersson & Potter 2010, Matsumura et al 2011). The specific, likely unique, prediction that the grain alignment should depend on the angle between the radiation and magnetic fields has been given support though our observations of the polarization surrounding HD 97300, in Chamaeleon.

If confirmed, the hinted at sharp drop-off in fractional polarization beyond $A_V \sim 10$ in starless cores would further support RAT alignment and would provide a unique tool to constrain the size distribution of the large grains. High sampling density polarimetry in environments with varying (but known!) radiation field SEDs (such as where dust can be proved next to very hot stars) could also probe the puzzling polarization behavious of the 2175Å extinction bump.

The lack of polarization of the carbonaceous grains, however, is a problem for RAT theory. If there are, indeed, large carbonaceous grain – as the extinction curve seems to require (Clayton et al. 2003, Draine & Fraisse 2009) – then why are these not aligned? Why would silicate grains exhibit helicity, but the carbonaceous grains not?

To paraphrase Hildebrand's 1988 review of this subject: While it is necessary to now set aside several of the proposed alignment mechanisms, it is probably not safe to dismiss them altogether. It is likely that each is important somewhere in the universe. It would, however, seem that - as far as the current observational data indicate - RAT theory is the best existing candidate to explain the general interstellar polarization.

**Acknowledgments** I am grateful to several colleagues for helpful discussions and for comments on early versions of the manuscript. I particular acknowledge many helpful conversations with and helpful suggestions by John Vaillancourt. In addition, Dan Clemens, Terry Jones, Alex Lazarian, Peter Martin and Doug Whittet generously responded to e-mail queries and read early frafts of the manuscript. Any errors or misstatements are, however, mine. The citations quoted in this review are by no means complete, given the long and expansive effort in the field. I apologize in advance to any of my colleagues who might feel slighted by missed or insufficiently strong citations and can only assure them that none of those oversights were intentional. I gratefully acknowledge financial support from the National Science Foundation, through grant AST-1109469